\documentclass[12pt,a4paper]{article}
\usepackage{amsmath,amsthm,amsfonts}
\usepackage{url}
\newcommand{\CC}{\mathbb{C}}
\newcommand{\RR}{\mathbb{R}}
\newcommand{\ZZ}{\mathbb{Z}}
\newcommand{\NN}{\mathbb{N}}

\newcommand{\LL}{\mathcal{L}}

\DeclareMathOperator{\spec}{spec}
\DeclareMathOperator{\mspec}{\mu{-}spec}

\newtheorem{prop}{Proposition}[section]
\newtheorem{theorem}[prop]{Theorem}
\newtheorem{corol}[prop]{Corollary}
\newtheorem{defin}[prop]{Definition}
\newtheorem{lem}[prop]{Lemma}
\newtheorem{remark}[prop]{Remark}

\numberwithin{equation}{section}

\renewcommand{\Hat}{\widehat}

\begin{document}

\title{\bf Semiclassical reduction\\
for magnetic Schr\"odinger operator\\
with periodic zero-range potentials\\
and applications
}

\author{\Large Bernard Helffer \dag  \and \Large Konstantin Pankrashkin \strut\dag\ddag\S\\[\bigskipamount]
\dag{} Universit\'e Paris Sud\\
D\'epartement de Math\'ematiques, B\^atiment 425\\
91405 Orsay Cedex, France\\[\medskipamount]
\ddag{} Humboldt-Universit\"at zu Berlin, Institut f\"ur Mathematik\\
Rudower Chaussee 25, 12489 Berlin, Germany\\[\medskipamount]
\S{} Corresponding author\\
E-mail: const@mathematik.hu-berlin.de\\
Phone +49 30 2093 2352, Fax +49 30 2093 2727}

\date{}

\maketitle

\begin{abstract} \noindent
The two-dimensional Schr\"odinger
operator with a uniform magnetic field
and a periodic zero-range potential is considered.
For weak magnetic fields and a weak coupling we reduce the spectral problem
to the semiclassical analysis of one-dimensional Harper-like operators.
This shows the existence of parts of Cantor structure in the spectrum
for special values of the magnetic flux.
\end{abstract}

{\bf Keywords.}
Schr\"odinger operator, periodic perturbation,
Cantor spectrum, magnetic field, zero-range potential

\newpage

\section{Introduction}

The spectral properties of a charged particle in a two-dimensional
system submitted to a periodic electric potential and a uniform
magnetic field crucially depend on
the arithmetic properties of 
the number $\theta$ representing  the magnetic flux quanta through the elementary cell
of periods, see e.g. \cite{bel} for a description of various models.
Since the works by Azbel \cite{AZ} and Hofstadter \cite{Hof}
it is generally believed that for irrational $\theta$ the spectrum
is a Cantor set, and the graphical presentation of the dependence of the spectrum on $\theta$
shows a fractal behavior known as the Hofstadter butterfly.
(We note that fractal spectral diagrams has been
found recently in the three-dimensional situation as well, see e.g. \cite{BDG}).
After  intensive efforts during more than twenty years 
 this was rigorously proved recently (Ten Martini conjecture) 
for all irrational values of $\theta$ for the discrete Hofstadter model, i.e.
the discrete magnetic Laplacian admitting a reduction to the almost Mathieu
equation, see \cite{AJ} and references therein.

Only few results are available for other models. Traditionally,
semiclassical methods have played an important role in the analysis of the two-dimensional
magnetic Schr\"odinger operators with periodic potentials, see e.g. \cite{BDP} for a review.
In particular, the bottom part of the spectrum for strong magnetic fields
can be described up to some extent using the tunneling asymptotics in a very general setting~\cite{HK}.
Concerning a more detailed analysis, it was shown by Helffer and Sj\"ostrand \cite{HS1,HS2,HS3} that the study of some parts of the spectrum
for the Schr\"odinger operator with a magnetic field and a periodic electric potentials
reduces to the spectral problem for an operator pencil of one-dimensional
quasiperiodic pseudodifferential operators (see below, Section~\ref{op1}); recently this correspondence
was extended up to a unitary equivalence \cite{Eck-cras,Eck}. Under some symmetry
conditions for the electric potentials, the operator pencil reduces to the study
of small perturbation of the continuous analog of the almost-Mathieu operator,
which allowed one to carry out a rather detailed iterative analysis for special values
of $\theta$. In particular, in several asymptotic regimes a Cantor structure
of some part of the  spectrum was proved.

In the present paper we are interested in the spectrum of the two-dimensional magnetic
Schr\"odinger operator with periodic \emph{zero-range} potentials (called also point perturbations).
Perturbations of such a kind became a rather popular model in quantum mechanics
due to a possibility of an analytic investigation whose results are in a good agreement
with experiments, see the monographs~\cite{AGHH,DO}.
Various aspects of the spectral analysis of magnetic Hamiltonians with point perturbations
were discussed in numerous works, see e.g. \cite{Avi1,Avi2,DMP,EJK,Gey,Gey2,Gre2,PS} and references therein.
Our principal aim in this paper is
to show that the technique of Helffer and Sj\"ostrand is still applicable
and is general enough to handle this non-standard class of operators;
such a correspondence between the semiclassical analysis and solvable models
seems to appear for the first time. This provides a proof of the existence
of Cantor parts in the spectrum in a certain asymptotic situation.

Our starting point will be the construction of \cite{Gey}
for the resolvent of the perturbed operator, which reduces the spectral problem
to the study of a family of discrete operators (see Section~\ref{sec1}).
We then construct a family of pseudodifferential operators (effective Hamiltonians)
having the same spectrum as the discrete operators (Section~\ref{sec3}).
In Section~\ref{sec4} we obtain some estimates for the Green function
of the Landau Hamiltonians, which are used to show that in the weak magnetic field
limit the effective Hamiltonians obey the conditions needed for the Helffer-Sj\"ostrand analysis (Section~\ref{sec5}
and the main result in Theorem \ref{th10}).

\section{Magnetic operator with zero-range potentials}\label{sec1}

We start with the two-dimensional Schr\"odinger operator with a uniform magnetic field
in the Landau gauge (Landau Hamiltonian),
\[
H_h= \Big(-i\dfrac{\partial}{\partial x_1}+h x_2\Big)^2-\dfrac{\partial^2}{\partial x_2^2}, \quad h>0.
\]
Recall that the spectrum of $H_h$ consists of the infinitely degenerate eigenvalues
$E_n=(2n-1)h$, $n\in\NN$, called the Landau levels.
We are going to study periodic zero-range perturbations of $H_h$ supported
by the set $\ZZ^2$. Physically, such operators model periodic
arrays of identical small impurities. The interactions can be intuitively
understood as operators corresponding to the limit for $\varepsilon\to0$
of the operators
\[
H(\varepsilon)=H_h+\alpha(\varepsilon)\sum_{m\in\ZZ^2}V_\varepsilon(\cdot-m),
\]
for localized potentials $V_\varepsilon$ approaching 
the Dirac $\delta$ function and a special choice
of the constants $\alpha(\varepsilon)$.
An alternative (but essentially equivalent) approach, which is more suitable for the spectral analysis,
involves the use of self-adjoint extensions \cite{BGP2}. In this case
one considers first the restriction of $H_h$ to the functions vanishing at all points of  $\ZZ^2$ (this operator is well defined as all functions
in the domain of $H_h$ are continuous), to be denoted by $S$.
By zero-range perturbations of $H_h$, one means then self-adjoint extensions of $S$.

We restrict ourselves by considering the standard one-parameter family of operators (self-adjoint extensions)
$H_{h,\alpha}$, $\alpha\in\RR\setminus\{0\}$, whose domain consists
of the functions $f$ having logarithmic singularities at the points of $\ZZ^2$,
\begin{gather*}
f(x)=\dfrac{a_m(f)}{2\pi}\log\dfrac{1}{|x-m|}+b_m(f)+o(1) \text{ as } |x-m|=o(1),\\
a_m(f), b_m(f)\in\CC,\quad m\in \ZZ^2, 
\end{gather*}
and satisfying the ``boundary conditions''
$a_m(f)+\alpha b_m(f)=0, \quad m\in\ZZ^2$. The action of the operators
in the sense of distributions
is given by the following expression (Fermi pseudopotential):
\[
H_{h,\alpha} f(x):=H_h f(x)+
\alpha\sum_{m\in\ZZ^2} \delta(x-m)\big(1+\log|x-m| \,\langle x-m,\nabla\rangle\big) f(x),
\]
where $\delta$ is the Dirac delta function.
The above boundary conditions satisfy the natural assumption
of form-locality \cite{KP01}, i.e.  the associated bilinear form vanishes on functions
 with disjoint supports.
This helps to avoid some pathological spectral properties appearing
for more general boundary conditions~\cite{GP}.

Let us repeat the  main constructions of \cite{Gey} to show how to handle the spectral problem for $H_{h,\alpha}$.
Note first that the Green function (the integral kernel of the resolvent) of $H_h$ is
known explicitly,
\begin{multline}
          \label{eq-green}
G_h(x,y;z)=\dfrac{1}{4\pi}\,\exp\Big(- \dfrac{i h (x_1-y_1)(x_2+y_2)}{2}-\dfrac{h(x-y)^2}{4}\Big)\\
\times \Gamma\Big(\dfrac{1}{2}-\dfrac{z}{2h}\Big) U\Big(\dfrac{1}{2}-\dfrac{z}{2h},1;\dfrac{h(x-y)^2}{2}\Big),
\end{multline}
where $U$ is the Kummer confluent hypergeometric function. The Green function
plays a crucial role in the spectral analysis of the operators
$H_{h,\alpha}$.
Namely, let us  define, for $z\notin\spec H_h$,
\begin{multline}
      \label{eq-qb}
q(z,h):=\lim_{x\to y} \Big(
G_h(x,y;z)-\dfrac{1}{2\pi}\log\dfrac{1}{|x-y|}
\Big)\\
=-\dfrac{1}{4\pi}\Big(
\psi\big(\dfrac{1}{2}-\dfrac{z}{2h}\big)- \log \dfrac{h}{2}-2\psi(1)
\Big),
\end{multline}
where $\psi$ is the logarithmic derivative of the $\Gamma$-function.
Note that the limit is independent of the choice of $y\in\RR^2$.
For $z\notin \spec H_h$ define an operator $Q(z,h):\ell^2(\ZZ^2)\to \ell^2(\ZZ^2)$ given
in the canonical basis by the matrix
\[
Q(m,n;z,h)=\begin{cases}
q(z,h), & m=n,\\
G_h(m,n;z),& \text{otherwise.}
\end{cases}
\]
Then the Green function $G_{h,\alpha}$ of $H_{h,\alpha}$ is given, for $z\notin\spec H_h\cup\spec H_{h,\alpha}$, by
\begin{multline}
  \label{eq-greenf}
G_{h,\alpha}(x,y;z)=G_h(x,y;z)\\
-\alpha\sum_{m,n\in\ZZ^2} \big(\alpha Q(z,h)+1\big)^{-1}(m,n)\, G_h(x,m;z)G_h(n,y;z),
\end{multline}
where the series on the right-hand side converges in the strong resolvent sense. 
An important ingredient of the above formula \eqref{eq-greenf} is the relation
\begin{equation}
       \label{eq-specba}
\spec H_{h,\alpha}\setminus \spec H_h=\big\{z\notin\spec H_h:\, 0\in\spec\big(Q(z,h)+\alpha^{-1}\big)\big\}.
\end{equation}
As the set $\spec H_h$ is discrete (the set of the Landau levels), Eq.~\eqref{eq-specba}
provides an almost complete characterization of the spectrum of $H_{h,\alpha}$,
and this is the starting point for the subsequent analysis.
Of crucial importance will be also the identity
\begin{equation}
      \label{eq-qinv}
Q(m+k,n+k;z,h)=e^{-ih k_2(m_1-n_1)} Q(m,n;z,h), \quad m,n,k\in\ZZ^2,
\end{equation}
which can be verified directly; this expresses the invariance of $H_h$ under the  magnetic translations \cite{Zak}.

\section{Strong type I operators}\label{op1}

In the works \cite{HS1,HS2,HS3} a machinery was developed
for an iterative semiclassical analysis of a special class of pseudodifferential operators.
One was concerned with the non-linear spectral problem
(or, in other words, with the spectral problem for an operator pencil).
Namely, for a family of self-adjoint operators $A(\mu)$
depending on a real parameter $\mu$, one means 
   by the $\mu$-spectrum $\mspec A(\mu)$
the set
of all $\mu$ such that $0\in\spec A(\mu)$. In particular,
the $\mu$-spectrum of the family $A(\mu):=A-\mu$ for a self-adjoint operator $A$
is exactly the spectrum of $A$.

Note that in the above terms the problem \eqref{eq-specba}
is exactly to find the $z$-spectrum for the family $Q(z,h)+\alpha^{-1}$.

Consider a bounded function (symbol) $L:\RR^2\to \CC$, $L=L(x,p)$.
By the Weyl quantization procedure
one can assign to $L$ an operator $\Hat L_h$ in $\LL^2(\RR)$,
\begin{equation}
      \label{eq-hweyl}
\Hat L_h f(x)=\dfrac{1}{2\pi h}\int_\RR \int_\RR
e^{ip(x-y)/h} L\big(\dfrac{x+y}{2},p\big) f(y) dp\,dy\,.
\end{equation}
Note that $L$ can depend itself on $h$ and other parameters as well.
The operator $\Hat L_h$ obtained is referred to as the Weyl $h$-quantization of $L$
(or the $h$-pseudodifferential operator with the symbol $L$),
and quantum Hamiltonians resulting from symbols periodic in both $x$ and $p$
are often called Harper-like operators.
In particular, the symbol $L(x,p):= \cos p +\cos x$ produces the Harper operator on the real line,
\begin{equation}
      \label{eq-harp}
\Hat L_h f(x) = \dfrac{f(x+h)+f(x-h)}{2}+\cos x\, f(x).
\end{equation}

In \cite{HS3}, in order to treat the Harper operator
and its  perturbations occurring
in a renormalization procedure,  the following notion was introduced,
see Definition 3.1 in \cite{HS3} with $L=P$ and $P_1=P_2=I$.
\begin{defin}\label{defin1}  A symbol $L(x,p;\mu,h)$ will be called of strong type~I if the following
conditions are satisfied for all $h\in(0,h_0)$ with some $h_0>0$\textup{:}
\begin{itemize}
\item[\rm(a)] $L$ depends analytically on $\mu\in[-4,4]$ and takes real values for such $\mu$.
\item[\rm(b)] There exists $\varepsilon>0$ such that
\begin{itemize}
\item[\rm(b1)] $L(x,p;\mu,h)$ is holomorphic in 
\[
D_\varepsilon=\Big\{(\mu,x,p)\in \CC \times\CC\times \CC:  |\mu|\le
4, \, |\Im x|+|\Im p|<\frac{1}{\varepsilon}\Big\},
\]
\item[\rm(b2)] for $(\mu,x,p)\in D_\varepsilon$, there holds
\[
\big|L(x,p;\mu,h)-(\cos x+\cos p+\mu)\big|\le\varepsilon.
\]
\end{itemize}
\item[\rm(c)] The following symmetry and periodicity conditions hold\textup{:}
\begin{itemize}
\item[\rm(c1)] $L(x,p;\mu,h)=L(-p,x;\mu,h)=L(x,-p;\mu,h)$,
\item[\rm(c2)] $L(x,p;\mu,h) =L(x+2\pi,p;\mu,h)= L(x,p+2\pi;\mu,h)$.
\end{itemize}
\end{itemize}
By $\varepsilon(L)$ we will denote the infimum of $\varepsilon$ for which the above conditions hold.
\end{defin}

In \cite{HS1,HS2,HS3} a detailed analysis was performed for pseudodifferential operators associated
with strong type~I symbols. One of the results (appearing as an
intermediate
 stronger result toward the proof of Theorem 0.1 in \cite{HS3})  was
\begin{theorem}\label{ths}
Let $L(\mu,h)$ be a strong type~I symbol. There exists $\varepsilon_0>0 $
and $C>0$ such that if  $\varepsilon(L)\le \varepsilon_0$ and 
if  $(2\pi)^{-1}h$ is an irrational admitting a representation
as a continuous fraction 
\[
\dfrac{h}{2\pi}=\dfrac{1}{n_1+\dfrac{\mathstrut 1}{n_2+\dfrac{\mathstrut 1}{n_3+\dots}}}
\]
with integers $n_j$ satisfying $n_j\ge C$, then the $\mu$-spectrum of the associated family of operators $\Hat L_h(\mu)$
is a zero measure Cantor set.
\end{theorem}
For operators $H=(-i\nabla+A)^2  +V$ with periodic potentials $V$,
\[
V(x_1+2\pi,x_2)\equiv V(x_1,x_2+2\pi) \equiv V(x_1,x_2),
\]
and constant magnetic fields, $\mathop{\rm curl} A = B$, 
it was shown in several asymptotic regimes that the study of some parts of the spectrum
reduces to a non-linear spectral problem of the above type for $B^{-1}$-pseudodifferential operators with symbols
close to $V(x,p)$ for strong magnetic fields, see e.g. \cite{HS4},
or for  $B$-pseudodifferential  operators with principal 
symbols coinciding with the first band function
of the zero-field Hamiltonian (Peierls substitution) in the weak magnetic field limit, see \cite[Appendix e]{HS1},
\cite[p.~117]{HS3}, \cite{HS4}, and earlier contributions by physicists mentioned e.g. in \cite{bel}. 
Hence, strong~type I operators appear when considering potentials 
 or  first band functions close
to $\beta(\cos x_1+\cos x_2) + \gamma $ for some reals $\beta\neq 0$,
$\gamma$.

Note that the expression \eqref{eq-greenf} is a realization of the Krein resolvent formula for self-adjoint extensions, see e.g. \cite{BGP2},
and can be viewed as a kind of the resolvent formula arising in the study of Grushin problems
\cite{SZ}, which suggests a certain analogy with the case of regular periodic perturbations studied in \cite{HS4}.
Our aim here is to show that the spectral problem \eqref{eq-specba} for periodic point perturbations
can be also reduced in a certain regime (weak magnetic field and weak coupling) to the study of pseudodifferential
operators with strong type I symbols.

\section{Symbols associated with some discrete operators}\label{sec3}

It is well known that the spectrum of the operator \eqref{eq-harp} as a set coincides with the spectrum
of the discrete magnetic Laplacian acting on $\ell^2(\ZZ^2)$, see e.g. \cite[pages 10--11]{HS1},
\[
C_h f(m,n)=e^{ih n}f(m+1,n)+e^{-ih n}f(m-1,n)+f(m,n-1)+f(m,n+1).
\]
The following theorem describes a similar correspondence for more
general operators. 
This is essentially a suitable reformulation of the constructions
of Section~6 in~\cite{HS4}.

\begin{theorem}\label{prop-red} Let $C_h$ be a bounded linear
operator on $\ell^2(\ZZ^2)$ given by
an infinite matrix $\big(C(p,q)\big)$, $p,q\in\ZZ^2$,
with exponentially decreasing entries,
$|C(p,q)|\le a e^{-b|p-q|}$ for some $a,b>0$ and all $p,q\in\ZZ^2$, and 
satisfying
\begin{equation}
     \label{eq-inv}
C(p+k,q+k)=e^{-i{h} k_2(p_1-q_1)} C(p,q),\quad p,q,k\in\ZZ^2,
\end{equation}
with some $h>0$.
Then the spectrum of $C_h$ coincides with the spectrum
of the Weyl $h$-quantization of the \textup{(}$h$-dependent\textup{)} symbol $T$ given by
\[
T(x,p)=\sum_{m,n\in\ZZ} c(m,n) e^{-imn h/2} e^{i(mx+np)},
\]
where $c(m,n)=C\big((0,0),(m,n)\big)$, $m,n\in\ZZ$.
\end{theorem}

\begin{proof}
Let us start with preliminary constructions.
Let $T:\RR^2\to \CC$ be a periodic smooth function,
$T(x,p+2\pi)=T(x+2\pi,p)=T(x,p)$.
Consider its Fourier expansion,
\begin{equation}
      \label{eq-hfourier}
\begin{gathered}
T(x,p)=\sum_{m,n\in\ZZ} t(m,n) e^{i(mx+np)},\\
t(m,n)=\dfrac{1}{(2\pi)^2}\int_0^{2\pi}\int_0^{2\pi} T(x,p) e^{-i(mx+np)}dp \,dx.
\end{gathered}
\end{equation}
We will assume that the coefficients $t(m,n)$ are exponentially decaying, i.e. that
\[
\big|t(m,n)\big|\le a e^{-b\sqrt{m^2+n^2}}\,,\,\mbox{  for some }
a,b>0\;.
\]
Let us find an expression for the corresponding operator $\Hat T_h$ given by \eqref{eq-hweyl}.
Let $f\in\mathcal{S}(\RR)$.
Substituting \eqref{eq-hfourier} into \eqref{eq-hweyl}
we obtain
\begin{multline}
   \label{eq-hw1}
\Hat T_h f(x)=\dfrac{1}{2\pi h}\sum_{m,n\in\ZZ} t(m,n)
\int_\RR \int_\RR e^{i(np+m(x+y)/2)} e^{ip(x-y)/h} f(y) dp\,dy\\
=\dfrac{1}{2\pi h}\sum_{m,n\in\ZZ} t(m,n) e^{imx/2}
\int_\RR \int_\RR e^{i(np+my/2)} e^{ip(x-y)/h} f(y) dp\,dy.
\end{multline}
Note that for each $x\in\RR$ and $n\in\ZZ$ there holds, in the sense of distributions,
\[
\int_\RR e^{ip(x-y)/h +inp} dp=-2\pi\delta\big(\dfrac{x-y}{h}+n\big).
\]
Hence  Eq. \eqref{eq-hw1} rewrites as
\begin{align}
\Hat T_h f(x)&=\dfrac{1}{h}\sum_{m,n\in\ZZ}t(m,n)e^{imx/2}
\int_\RR  e^{imy/2}  \delta(\dfrac{x-y}{h}+n) f(y)  dy\notag\\
&=\sum_{m,n\in\ZZ} t(m,n)e^{imx/2} \Big(e^{imy/2} f(y)\Big)_{y=x+nh}\notag\\
   \label{eq-hw2}
&=\sum_{m,n\in\ZZ} t(m,n)e^{imx} e^{imnh/2} f(x+nh),
\end{align}
and due to the exponential decay of the coefficients $t(m,n)$
this extends by continuity to the whole of $\LL^2(\RR)$.

Let us return to the initial operator $C_h$. By the assumption \eqref{eq-inv},
\[
C(p,q)=\exp\big(i{h} p_2(q_1-p_1)\big) c(q-p)
\mbox{ for any } p,q\in\ZZ^2\;.
\]
 Hence
\[
C_hf(p)=\sum_{q\in\ZZ^2} e^{i{h} p_2(q_1-p_1)} c(q-p) f(q)\\
=\sum_{q\in\ZZ^2} e^{i{h} p_2 q_1} c(q) f(p+q).
\]
Therefore, $C_h$ commutes with the shift $f(p_1,p_2)\mapsto f(p_1+1,p_2)$,
and the Floquet-Bloch theory is applicable.

Let us introduce the functions
\[
\RR\ni \varphi \mapsto b_n(\varphi)=\sum_{k\in\ZZ} c(k,n)e^{ik\varphi}, \quad n\in\ZZ.
\]
All these functions are $2\pi$-periodic  and analytic
in a complex neighborhood of $\RR$. Consider a family (indexed by
$\theta\in \mathbb R$) of operators $C_h(\theta)$
acting in 
$\ell^2(\ZZ)$,
\begin{equation*}
C_h(\theta) g(m)=\sum_{n\in\ZZ} b_n(m{h}+\theta) g(m+n), \quad m\in\ZZ\;. 
\end{equation*}
The operator-valued function $\theta\mapsto C_h(\theta)$ is obviously
continuous  in the norm topology.
Due to periodicity of the functions $b_n$ one has 
\begin{equation*}
C_h(\theta)=C_h(\theta+2\pi)\;.
\end{equation*}
Therefore, by the Floquet-Bloch theory, one has
\begin{equation}\label{invariance}
 \spec C_h=\bigcup_{\theta\in[0,2\pi)} \spec C_h(\theta)\;.
\end{equation}
Furthermore, for any $\theta$ the operators $C_h(\theta)$ and $C_h(\theta+{h})$ are unitarily equivalent,
\[
C_h(\theta+{h})=S C_h(\theta) S^{-1}\;,
\]
 where $S$ is the shift in $\ell^2(\ZZ)$, $$S f(n)=f(n+1)\;.$$
This implies, cf. \eqref{invariance}, \[ \spec C_h=\bigcup_{\theta\in[0,h)} \spec
 C_h(\theta)\;,
\]
 which
coincides with the spectrum of the following
operator $T_h$ acting in $\LL^2\big(\ZZ\times[0,{h})\big)$:
\[
T_h u(m,\theta)=C_h(\theta) u_\theta(m), \quad u_\theta(m)=u(m,\theta), \quad m\in\ZZ.
\]
Consider the map $U:\LL^2(\RR)\to \LL^2\big(\ZZ\times[0,{h})\big)$ given by
$$ U f(m,\theta)= f(m{h}+\theta)\;,\;(m,\theta)\in\ZZ\times[0,{h})\;.
$$
One can easily see that $U$ is unitary.
The operator $\Hat T_h:=U^{-1} T_h U$ is then an operator in $\LL^2(\RR)$,
\[
\Hat T_h f(x)=\sum_{n\in\ZZ} b_n(x) f(x+n{h})
=\sum_{m,n\in\ZZ} c(m,n)e^{im x} f(x+n{h}).
\]
Comparing this expression with \eqref{eq-hw2} we arrive at the conclusion.
\end{proof}

Using Theorem~\ref{prop-red} and the relations \eqref{eq-specba} and \eqref{eq-qinv} 
one arrives at the following corollary. 
\begin{corol}\label{corol2}
The set $\spec H_{h,\alpha}\setminus \spec H_h$ coincides with the $z$-spectrum
of the family of $h$-pseudodifferential operators
corresponding to the symbols
\begin{align}
      \label{eq-LL}
L_\alpha(x,p;z,h)&=\sum_{m,n\in\ZZ}Q\big((0,0),(m,n);z\big)e^{-imnh/2} e^{i(mx+np)}+\alpha^{-1}\\
&=q(z,h)+\alpha^{-1}+\sum_{\substack{m,n\in\ZZ,\\|m|+|n|\ge 1}}\lambda(m,n;z,h) e^{i(mx+np)}\notag
\end{align}
where
\begin{multline}
       \label{eq-lambda}
\lambda(m,n;z,h)=F_h\big((0,0),(m,n);z\big)\\
{}=\dfrac{1}{4\pi}\,\exp\Big(-\dfrac{h(m^2+n^2)}{4}\Big) \Gamma\Big(\dfrac{1}{2}-\dfrac{z}{2h}\Big) U\Big(\dfrac{1}{2}-\dfrac{z}{2h},1;\dfrac{h(m^2+n^2)}{2}\Big),
\end{multline}
and 
\begin{equation}\label{defFh}
F_h(x,y;z)=\exp\Big(\dfrac{i h (x_1-y_1)(x_2+y_2)}{2}\Big)\,G_h(x,y;z).
\end{equation}
\end{corol}
The operators $\Hat L_{h,\alpha}$ associated with the above symbols $L_\alpha$ can be viewed as \emph{effective Hamiltonians}
of the problem; we emphasize that these are  given in an explicit form (the Fourier expansion).
We are going to show that after suitable transformations one arrives
at the study of strong type I symbols.

\section{Estimates for the Green function}\label{sec4}

In the present section we will establish some estimates involving the Green function, the spectral parameter,
and the strength of the magnetic field in the form which will be of importance below. The representation \eqref{eq-green}
works effectively near the Landau levels, see for example \cite{EP},
 while we intend to consider the negative halfline. So we prefer to use
the corresponding heat kernel, i.e. the integral kernel of the semigroup $e^{-t H_h}$, $t>0$, see e.g. Section 6.2.1.5 in~\cite{GS},
\begin{multline}
   \label{eq-heat}
P_h(x,y;t)=\dfrac{h}{4\pi \sinh ht}\,
\exp\Big(- \dfrac{i h (x_1-y_1)(x_2+y_2)}{2}\Big)\\
\times  \exp\Big(-\dfrac{h(x-y)^2 \coth ht}{4}\Big).
\end{multline}
To unify the notation, we denote  by $H_0$ the free Laplacian (i.e., the operator $H_h$ with $h=0$).
The above expression \eqref{eq-heat} will be used for $h>0$, and by $P_0$ we denote the heat kernel associated
with the free Laplacian,
\[
P_0(x,y;t)=\dfrac{1}{4\pi t} \exp\Big(-\dfrac{(x-y)^2}{4t}\Big).
\]
For the Green functions we will use the integral representation
\begin{equation}
     \label{eq-int}
G_h(x,y;z)=\int_0^\infty e^{zt}P_h(x,y;t)dt,\quad h\ge 0,
\end{equation}
which is valid at least for $\Re z<h$. For the function $F_h$ introduced in \eqref{defFh} 
one has
\begin{equation}
     \label{eq-intf}
F_h(x,y;z)=\int_0^\infty e^{zt}\big|P_h(x,y;t)\big|dt,\quad h>0.
\end{equation}

Recall that the Green function $G_0$
of the free Laplacian is known explicitly as well,
\begin{equation}
   \label{eq-gf0}
G_0(x,y;z)=\dfrac{1}{2\pi}\, K_0\big(\sqrt{-z}\,|x-y|\big),
\end{equation}
where $K_0$ is the modified Bessel function of order zero.
Here and below we fix the branch of the square root which
is a continuation of the usual square root on the positive half-line.

We will also use the integral kernel $G^{(2)}_h(x,y;z)$ of the operator $(H_h-z)^{-2}$, $h\ge 0$, $\Re z<0$,
and the function
\[
F^{(2)}_h(x,y;z)=\exp\Big(\dfrac{i h (x_1-y_1)(x_2+y_2)}{2}\Big)\,G^{(2)}_h(x,y;z)\;.
\]
By the Hilbert resolvent identity, there holds
\begin{equation}
     \label{eq-gh2}
\dfrac{\partial G_h(x,y;z)}{\partial z}=G^{(2)}_h(x,y;z),
\end{equation}
and, at the same time,
\begin{equation}
     \label{eq-fh2}
\dfrac{\partial F_h(x,y;z)}{\partial z}=F^{(2)}_h(x,y;z).
\end{equation}

Using the identity $K'_0(w)=-K_1(w)$, see Eq.~(9.6.27) in~\cite{AS}, and Eq. \eqref{eq-gf0} above, we obtain
\begin{equation}
      \label{eq-g20}
G^{(2)}_0(x,y;z)=\dfrac{|x-y|}{4\pi\sqrt{-z}}\, K_1\big(\sqrt{- z}\,|x-y|\big).
\end{equation}
On the other hand, using
\[
(H_h-z)^{-2}=\int_0^\infty t e^{-t(H_h-z)}dt, \quad  \Re z <0, \quad h\ge 0
\]
we arrive at 
\begin{gather}
      \label{eq-g2int}
G^{(2)}_h(x,y;z)= \int_0^\infty t e^{zt} P_h(x,y;t)\,dt, \quad  \Re z <h, \quad h\ge 0,\\
      \label{eq-f2int}
F^{(2)}_h(x,y;z)= \int_0^\infty t e^{zt} \big|P_h(x,y;t)\big|\,dt, \quad  \Re z <h, \quad h>0.
\end{gather}

The following two-side estimate will be of importance.
\begin{lem}
    \label{lem-heat}
For  $h>0$ there holds
\[
\exp\Big(-ht-\dfrac{h(x-y)^2}{4}\Big)P_0(x,y;t)
\le \big|P_h(x,y;t)\big|\le P_0(x,y;t)
\]
for any $t>0$ and $x,y\in\RR^2$.
\end{lem}

\begin{proof} The right-hand side inequality  is nothing but the
Kato (diamagnetic) inequality valid for much more general magnetic fields, see e.g. \S{2} in \cite{AHS}, so we only need to prove the left-hand
side inequality. We will use the elementary estimates
\begin{equation}
      \label{eq-elem}
      \dfrac{e^t-1}{t}\ge 1 \text{ and }
\dfrac{t}{1-e^{-t}}\ge 1 \text{ for } t>0.
\end{equation}
Using \eqref{eq-elem} one has
\[
\dfrac{h}{\sinh ht}=\dfrac{e^{-ht}}{t}\,\dfrac{2ht}{1-e^{-2ht}}\ge \dfrac{e^{-ht}}{t}.
\]
At the same time, using \eqref{eq-elem} again, one obtains
\[
h \coth ht=h\Big(1+\dfrac{2}{e^{2ht}-1}\Big)= h+ \dfrac{1}{t}\, \dfrac{2ht}{e^{2ht}-1}\le h +\dfrac{1}{t}.
\]
Therefore,
\begin{align*}
\big|P_h(x,y;t)\big|&=\dfrac{h}{4\pi \sinh ht}\,  \exp\Big(-\dfrac{h(x-y)^2 \coth ht}{4}\Big)\\
&\ge \dfrac{1}{4\pi t} \exp\Big(
-ht - \dfrac{h(x-y)^2}{4}-\dfrac{(x-y)^2}{4t}
\Big)\\&=\exp\Big(
-ht - \dfrac{h(x-y)^2}{4}\Big) P_0(x,y;t). \qedhere
\end{align*}
\end{proof}

Using \eqref{eq-int} and \eqref{eq-g2int} one arrives at the following corollary.
\begin{corol}\label{prop-zb}
For $x,y\in\RR^2$, $x\ne y$, and real $z$, $z<h$, one has
\[
\big|G_h(x,y;z)\big|\ge G_0(x,y;z-h) \exp\Big(- \dfrac{h(x-y)^2}{4}\Big)
\]
and, for any $x,y\in\RR^2$ and real $z$ with $z<h$, there holds
\[
\big|G_h(x,y;z)\big|\ge G_0^{(2)}(x,y;z-h) \exp\Big(- \dfrac{h(x-y)^2}{4}\Big).
\]
\end{corol}

\begin{lem}\label{lem6} For $h>0$ and $\Re z<0$ there holds
\begin{multline}
  \label{eq-gest1}
\big|
F_h(x,y;z)-G_0(x,y;z)
\big|\\
\le  G_0(x,y;\Re z)-G_0(x,y;\Re z-h) \exp\Big(- \dfrac{h(x-y)^2}{4}\Big)
\end{multline}
and
\begin{multline}
  \label{eq-gest2}
\big|F^{(2)}_h(x,y;z)-G^{(2)}_0(x,y;z) \big|\\ \le  G^{(2)}_0(x,y;\Re z)-G^{(2)}_0(x,y;\Re z-h) \exp\Big(- \dfrac{h(x-y)^2}{4}\Big).
\end{multline}
\end{lem}
\begin{proof}
Using \eqref{eq-int} and \eqref{eq-intf} one has
\begin{multline*}
|F_h(x,y;z)-G_0(x,y;z)\big|=\Big|\int_0^\infty e^{zt}
\big(\big|P_h(x,y;t)\big|-P_0(x,y;t)\big)dt\Big|\\
\le\int_0^\infty \big|e^{zt}\big|\cdot \Big|\big|P_h(x,y;t)\big|-P_0(x,y;t)\Big|\,dt.
\end{multline*}
Estimate $\big|e^{zt}\big|=e^{\Re z t}$ and, using Lemma \ref{lem-heat},
\begin{multline*}
\Big|\big|P_h(x,y;t)\big|-P_0(x,y;t)\Big|= P_0(x,y;t)-\big|P_h(x,y;t)\big|\\
\le P_0(x,y;t) - \exp\Big(-ht-\dfrac{h(x-y)^2}{4}\Big)P_0(x,y;t).
\end{multline*}
We obtain, using \eqref{eq-int} again,
\begin{multline*}
|F_h(x,y;z)-G_0(x,y;z)\big|\\
\begin{aligned}
~&
\le
\int_0^\infty e^{\Re z t} \Big(P_0(x,y;t) - \exp\Big(-ht-\dfrac{h(x-y)^2}{4}\Big)P_0(x,y;t)\Big)dt\\
&=\int_0^\infty e^{\Re z t} P_0(x,y;t)dt
 - \exp\Big(-\dfrac{h(x-y)^2}{4}\Big)\,\int_0^\infty e^{(\Re z-h) t} P_0(x,y;t)\,dt\\
&{}=G_0(x,y;\Re z)-G_0(x,y;\Re z-h) \exp\Big(- \dfrac{h(x-y)^2}{4}\Big).
\end{aligned}
\end{multline*}
Eq. \eqref{eq-gest2} is obtained in the same way using \eqref{eq-g2int} and \eqref{eq-f2int}.
\end{proof}

In what follows we need the function
\begin{equation}
       \label{eq-q0}
q_0(z):=\lim_{x\to y} \Big(
G_0(x,y;z)-\dfrac{1}{2\pi}\log\dfrac{1}{|x-y|}\Big).
\end{equation}
Using Eq. (9.6.13) in \cite{AS}, 
\begin{equation*}\label{AsympK0}
\lim_{r\to 0+} \big(K_0(w r)+\log r\big)= -\log w+\log 2+\psi(1), \quad w\in \CC\setminus (-\infty,0],
\end{equation*}
we obtain
\begin{equation}
\label{eq-qex}
q_0(z)=-\dfrac{\log(-z)-\log 4-2\psi(1)}{4\pi}.
\end{equation}
Everywhere we take the principal branch of the logarithm, i.e. the holomorphic extension
 to $\mathbb C \setminus (-\infty,0]$ of the usual logarithm on  $(0,+\infty)$.

\begin{lem}\label{Lemma8}
 For $h>0$ and $\Re z<0$ there holds
\begin{gather}
     \label{eq-qh1}
\big|q(z,h)-q_0(z)\big|\le \dfrac{1}{4\pi} \log\dfrac{\Re z-h}{\Re z},\\
     \label{eq-qh2}
\Big|\dfrac{\partial q(z,h)}{\partial z}-\dfrac{\partial q_0(z)}{\partial z}\Big|\le \dfrac{h}{4\pi \Re z(\Re z-h)}.
\end{gather}
\end{lem}

\begin{proof}
Consider first \eqref{eq-qh1}. Due to the logarithmic on-diagonal singularity of the Green functions
one has
\begin{multline*}
\lim_{x\to y} \Big(G_h(x,y;z)-F_h(x,y;z)\Big)\\
=\lim_{x\to y} 
\Big(1-\exp\big(\dfrac{i h (x_1-y_1)(x_2+y_2)}{2}\big)\Big)\,G_h(x,y;z)
=0.
\end{multline*}
Hence, using \eqref{eq-qb},
\begin{equation}
      \label{eq-qbh}
      q(z,h)=\lim_{x\to y} \Big(
F_h(x,y;z)-\dfrac{1}{2\pi}\log\dfrac{1}{|x-y|}\Big).
\end{equation}

Substituting now \eqref{eq-qbh} and \eqref{eq-q0} into \eqref{eq-gest1}
we see
\begin{multline*}
\big|q(z,h)-q_0(z)\big|=\lim_{x\to y} \Big|F_h(x,y;z)-G_0(x,y;z)\Big|\\
\le \lim_{x\to y}\Big(
G_0(x,y;\Re z)-G_0(x,y;\Re z-h) \exp\big(- \dfrac{h(x-y)^2}{4}\big)\Big)\\
= \lim_{x\to y} \Big( G_0(x,y;\Re z)-\dfrac{1}{2\pi}\log\dfrac{1}{|x-y|}\Big)\\
-\lim_{x\to y} \Big( G_0(x,y;\Re z-h)-\dfrac{1}{2\pi}\log\dfrac{1}{|x-y|}\Big)\exp\big(- \dfrac{h(x-y)^2}{4}\big)\\
+\lim_{x\to y} \dfrac{1}{2\pi}\, \Big(1-\exp\big(- \dfrac{h(x-y)^2}{4}\big) \Big)\log\dfrac{1}{|x-y|}
= q_0(\Re z)-q_0(\Re z-h),
\end{multline*}
and now one can use the explicit expression \eqref{eq-qex}.

As for \eqref{eq-qh2}, the expressions \eqref{eq-qb} for $q(z,h)$ and \eqref{eq-q0} for $q_0(z)$ together
with the analyticity of $G_h$ in $z$ ($h\ge 0$) imply
\[
\dfrac{\partial q(z,h)}{\partial z}=G^{(2)}_h(x,x;z)=F^{(2)}_h(x,x;z), \quad
\dfrac{\partial q_0(z)}{\partial z}=G^{(2)}_0(x,x;z),
\]
where $x\in\RR^2$ is arbitrary. Therefore, using \eqref{eq-gest2}, one obtains
\[
\Big|\dfrac{\partial q(z,h)}{\partial z}-\dfrac{\partial q_0(z)}{\partial z}\Big|
\le G^{(2)}_0(x,x;\Re z)- G_0^{(2)}(x,x;\Re z-h)=\dfrac{\partial q_0(\lambda)}{\partial \lambda}\Big|_{\lambda=\Re z-h}^{\lambda=\Re z},
\]
and, by \eqref{eq-qex}, one has
\[
\dfrac{\partial q_0(\lambda)}{\partial \lambda}=-\dfrac{1}{4\pi \lambda}
\]
which finishes the proof.
\end{proof}

Below we will use intensively the following
well-known asymptotics, Eq. (9.7.2) in \cite{AS}:
\begin{equation}
     \label{eq-ask}
K_0(w)=\sqrt{\dfrac{\pi}{2w}}\, e^{-w}\,\big(1+O(w^{-1})\big), \quad
|w| \to \infty,\quad |\arg w|\le\dfrac{\pi}{4}.
\end{equation}

Let us prove first several estimates from above for the Green function and its derivative.

\begin{lem}\label{lem-oben} For any $d,\delta>0$ there exists $E>0$ such that for $|x-y|> d$, $\Re z<-E$,
and $h\ge0$ there holds
\[
\big|G_h(x,y;z)\big|\le \dfrac{1+\delta}{2\pi}\sqrt[4]{\dfrac{\pi^2}{-4\Re z\,|x-y|^2}}\, e^{-\sqrt{-\Re z}\,|x-y|}.
\]
\end{lem}

\begin{proof}
Using \eqref{eq-int} and the right-hand side inequality of Lemma \ref{lem-heat},
for $\Re z<0$ we obtain
\begin{multline*}
\big|G_h(x,y;z)\big|=\Big|\int_0^\infty  e^{zt} P_h(x,y;t)\,dt\Big|\\
\le\int_0^\infty \big|e^{zt}\big|\cdot \big| P_h(x,y;t)\big|\,dt
\le \int_0^\infty e^{\Re z\, t} P_0(x,y;t) dt=G_0(x,y;\Re z).
\end{multline*}
Substituting here \eqref{eq-gf0} one obtains
\[
\big|G_h(x,y;z)\big|\le \dfrac{1}{2\pi}\, K_0(\sqrt{-\Re z}\,|x-y|),
\]
and the asymptotics \eqref{eq-ask} gives the result.

\end{proof}

\begin{lem}\label{lem-afh} For any $d,\delta>0$ there exists $E>0$ such that for $|x-y|>d$, $\Re z<-E$,
and $h\ge0$ there holds
\[
\Big|
\dfrac{\partial G_h(x,y;z)}{\partial z}
\Big|\le \dfrac{(1+\delta)}{4\pi} \sqrt[4]{\dfrac{\pi^2 |x-y|^2}{-4(\Re z)^3}}\, \exp\big({}-\sqrt{-\Re z}\,|x-y|\big).
\]
\end{lem}

\begin{proof}
Using Eqs. \eqref{eq-gh2}, \eqref{eq-g2int}, and the right-hand side inequality of Lemma~\ref{lem-heat},
for $\Re z<0$ we obtain
\begin{multline*}
\Big|\dfrac{\partial G_h(x,y;z)}{\partial z}\Big|=\big|G^{(2)}_h(x,y;z)\big|
\le\int_0^\infty t \big|e^{zt}\big|\cdot \big| P_h(x,y;t)\big|\,dt\\
\le \int_0^\infty t e^{\Re z\,t} P_0(x,y;t) dt=G^{(2)}_0(x,y;\Re z).
\end{multline*}
Hence, using \eqref{eq-g20} we arrive at
\[
\Big|\dfrac{\partial G_h(x,y;z)}{\partial z}\Big|\le \dfrac{|x-y|}{4\pi \sqrt{-\Re z}}\, K_1(\sqrt{-\Re z}\,|x-y|),
\]
and it is sufficient to use the asymptotics
\[
K_1(w)=\sqrt{\dfrac{\pi}{2w}}\, e^{-w}\,\big(1+O(w^{-1})\big), \quad |w|
\to\infty,
\quad |\arg w|\le\dfrac{\pi}{4},
\]
see Eq. (9.7.2) in \cite{AS}.
\end{proof}

Let us pass to estimates from below, which are of crucial importance.

\begin{lem}\label{lem-unten} For any $x,y\in\RR^2$, $x\ne y$, and any $c\in(0,1)$
there exist $E,\gamma,h_0>0$ \textup{(}which depend on $x,y,c$\textup{)} such that
\[
\big|
G_h(x,y;z)
\big|\ge \dfrac{1-c}{2\pi}\sqrt[4]{\dfrac{\pi^2}{-4\Re z\,|x-y|^2}}\, e^{-\sqrt{-\Re z}\,|x-y|}
\]
for $\Re z<-E$, $|\Im z|<\gamma$, and $h\in(0,h_0)$.
\end{lem}

\begin{proof}
By Lemma \ref{lem6},
\begin{multline}
        \label{eq-gstart}
\big|G_h(x,y;z)\big|=\big|F_h(x,y;z)\big|\\
\begin{aligned}
~&\ge \big|G_0(x,y;z)\big| - \big|F_h(x,y;z)-G_0(x,y;z)\big|\\
&\ge \big|G_0(x,y;z)\big| +G_0(x,y;\Re z-h) \exp\Big(- \dfrac{h(x-y)^2}{4}\Big) - G_0(x,y;\Re z)\\
&= \big|G_0(x,y;z)\big| +G_0(x,y;\Re z-h) \,\Big( \exp\Big(- \dfrac{h(x-y)^2}{4}\Big)-1\Big)\\
&\quad{}+ \big(G_0(x,y;\Re z-h)- G_0(x,y;\Re z)\big).
\end{aligned}
\end{multline}
Due to \eqref{eq-ask} and Lemma \ref{lem-oben},
for any $\delta\in(0,1)$ one can choose $E(\delta,x,y)>0$ 
such that
\begin{multline}
        \label{eq-gloc2}
\dfrac{1-\delta}{2\pi} \Big|\sqrt[4]{\dfrac{\pi^2}{-4 z|x-y|^2}}\, e^{-\sqrt{- z}|x-y|}\Big|\le
\big|G_0(x,y; z)\big|\\
\le \dfrac{1+\delta}{2\pi} \sqrt[4]{\dfrac{\pi^2}{-4 \Re z|x-y|^2}}\, e^{-\sqrt{- \Re z}|x-y|},\quad \Re z<-E(\delta,x,y)\,.
\end{multline}
The constant $\delta$ will be chosen later.

One can find $h_0=h_0(\delta,x,y)$ such that
\begin{equation}
     \label{eq-hhh0}
\Big| \exp\Big(- \dfrac{h_0(x-y)^2}{4}\Big)-1\Big|\le \delta,
\end{equation}
then, by the right-hand side inequality in \eqref{eq-gloc2},
\begin{multline}
     \label{eq-unt0}
\Big|G_0(x,y;\Re z-h) \,\Big( \exp\Big(- \dfrac{h(x-y)^2}{4}\Big)-1\Big)\Big|\\
\le \dfrac{\delta(1+\delta)}{2\pi}\sqrt[4]{\dfrac{\pi^2}{4(h-\Re z)|x-y|^2}}\, e^{-\sqrt{h- \Re z}|x-y|}\\
\le \dfrac{\delta(1+\delta)}{2\pi} \sqrt[4]{\dfrac{\pi^2}{-4\Re z\, |x-y|^2}}\, e^{-\sqrt{- \Re z}|x-y|}\\
\text{for } \Re z<-E(\delta,x,y), \quad h\in\big(0,h_0(\delta,x,y)\big).
\end{multline}

At the same time one has
\begin{multline*}
\Big|\dfrac{\exp\big(-\sqrt{-z}|x-y|\big)}{\sqrt[4]{-z}}\Big|=\dfrac{\exp\big(-\Re\sqrt{-z}|x-y|\big)}{\sqrt[4]{|z|}}\\
\ge\dfrac{\exp\big(-|\sqrt{-z}|\,|x-y|\big)}{\sqrt[4]{|z|}}
=\dfrac{\exp\big(-\sqrt{|z|}\,|x-y|\big)}{\sqrt[4]{|z|}}\\
\ge \dfrac{1}{\sqrt[4]{|\Re z|+|\Im z|}}\,\exp\big(-\sqrt{|\Re z|+|\Im z|}|x-y|\big),
\end{multline*}
hence for $|\Im z|<\gamma$ one can estimate
\begin{equation}
      \label{eq-low4}
\Big|\dfrac{1}{\sqrt[4]{-z}}\,\exp\big(-\sqrt{-z}\,|x-y|\big)\Big|
\ge \dfrac{1}{\sqrt[4]{\gamma+|\Re z|}}\,\exp\big(-\sqrt{\gamma+|\Re z|}|x-y|\big).
\end{equation}
For $ t>E>0$ and $\gamma>0$ one can write
\[
\sqrt{t+\gamma}-\sqrt{t}=\dfrac{\gamma}{\sqrt{t+\gamma}+\sqrt{t}}
\le \dfrac{\gamma}{2\sqrt{t}}\le \dfrac{\gamma}{2\sqrt{E}}\,.
\]
Hence, for any $\alpha,\gamma>0$ and $t>E>0$, one has
\begin{equation}
     \label{eq-elem4}
\exp\big(-\alpha\sqrt{t}\big)\le \exp\Big(\dfrac{\alpha\gamma}{2\sqrt E}\Big)\, \exp\big(-\alpha\sqrt{t+\gamma}\big).
\end{equation}
Analogously  one has for $t>E>0$ and $\gamma>0$, 
\begin{equation}
     \label{eq-elem5}
\sqrt[4]{\gamma+t}=\sqrt[4]{1+\dfrac{\gamma}{t}}\,\sqrt[4]{t}
\le \Big(1+\dfrac{\gamma}{t}\Big)\,\sqrt[4]{t}\le  \Big(1+\dfrac{\gamma}{E}\Big)\,\sqrt[4]{t}.
\end{equation}
Hence, assuming that $\gamma=\gamma(\delta,x,y)$ satisfies
\[
\exp \dfrac{\gamma|x-y|}{2\sqrt{E}}\le 1+\delta, \quad \dfrac{\gamma}{E}\le\delta,
\]
and substituting \eqref{eq-low4}, \eqref{eq-elem4}, and \eqref{eq-elem5} into
the left-hand side inequality in \eqref{eq-gloc2} one arrives at
\begin{multline}
          \label{eq-unt}
\big|G_0(x,y; z)\big|\ge \dfrac{1-\delta}{2\pi(1+\delta)^2} 
\sqrt[4]{\dfrac{\pi^2}{-4\Re z\,|x-y|^2}}\,  e^{-\sqrt{- \Re z}|x-y|}\\
\text{for }  \Re z<-E(\delta,x,y) \text { and } |\Im z|<\gamma(\delta,x,y).
\end{multline}

The estimates \eqref{eq-gloc2} give also, for $ \Re z<-E$,
\begin{multline*}
\big|G_0(x,y;\Re z-h)- G_0(x,y;\Re z)\big|\\
\begin{aligned}
~&\le
\dfrac{1+\delta}{2\pi} \sqrt[4]{\dfrac{\pi^2}{-4\Re z\,|x-y|^2}}\, e^{-\sqrt{-\Re z}|x-y|}\\
&\qquad -
\dfrac{1-\delta}{2\pi} \sqrt[4]{\dfrac{\pi^2}{4(h-\Re z)\,|x-y|^2}}\, e^{-\sqrt{h-\Re z}|x-y|}\\
&\le
\dfrac{1}{2\pi} \Big(\sqrt[4]{\dfrac{\pi^2}{-4\Re z\,|x-y|^2}} e^{-\sqrt{-\Re z}|x-y|}-
\sqrt[4]{\dfrac{\pi^2}{4(h-\Re z)\,|x-y|^2}} e^{-\sqrt{h-\Re z}|x-y|}\Big)\\
&\quad+ \dfrac{\delta}{\pi} \sqrt[4]{\dfrac{\pi^2}{-4\Re z\,|x-y|^2}}\, e^{-\sqrt{-\Re z}|x-y|}.
\end{aligned}
\end{multline*}

Expand
\begin{multline*}
\sqrt[4]{\dfrac{\pi^2}{-4\Re z\,|x-y|^2}}\, e^{-\sqrt{-\Re z}|x-y|}-
\sqrt[4]{\dfrac{\pi^2}{4(h-\Re z)\,|x-y|^2}}\, e^{-\sqrt{h-\Re z}|x-y|}\\
= \sqrt[4]{\dfrac{\pi^2}{-4\Re z\,|x-y|^2}} \,\Big(
e^{-\sqrt{-\Re z}|x-y|}-e^{-\sqrt{h-\Re z}|x-y|} \Big)\\
{}+ \bigg(
\sqrt[4]{\dfrac{\pi^2}{-4\Re z\,|x-y|^2}} -
\sqrt[4]{\dfrac{\pi^2}{4(h-\Re z)\,|x-y|^2}}
\bigg)\,e^{-\sqrt{h-\Re z}|x-y|}.
\end{multline*}
Using \eqref{eq-elem4} we estimate, for $h\in(0,h_0)$ and $\Re z<-E$,
\begin{multline*}
\exp\big(-\sqrt{-\Re z}|x-y|\big)-\exp\big(-\sqrt{h-\Re z}|x-y|\big)\\ 
\le \Big(\exp \dfrac{h|x-y|}{2\sqrt{E}}-1 \Big)\,\exp\big(-\sqrt{h-\Re z}|x-y|\big)\\
\le \Big(\exp \dfrac{h_0|x-y|}{2\sqrt{E}}-1 \Big)\,\exp\big(-\sqrt{-\Re z}|x-y|\big).
\end{multline*}
At the same time one has, for $h\in(0,h_0)$ and $\Re z<-E$,
\begin{multline*}
\sqrt[4]{\dfrac{\pi^2}{-4\Re z\,|x-y|^2}} -
\sqrt[4]{\dfrac{\pi^2}{4(h-\Re z)\,|x-y|^2}}\\
=\Big(\sqrt[4]{\dfrac{h-\Re z}{-\Re z}}-1\Big)\,
\sqrt[4]{\dfrac{\pi^2}{4(h-\Re z)\,|x-y|^2}}\\
\le \dfrac{h}{|\Re z|}\,\sqrt[4]{\dfrac{\pi^2}{4|\Re z|\,|x-y|^2}}
\le \dfrac{h_0}{E}\,\sqrt[4]{\dfrac{\pi^2}{4|\Re z|\,|x-y|^2}}.
\end{multline*}
Let us choose $h_0=h_0(\delta,x,y)$ in such a way that \eqref{eq-hhh0} still holds and that
\[
\exp\Big(\dfrac{h_0|x-y|}{2\sqrt{E}}\Big)-1<\delta, \quad \dfrac{h_0}{E}<\delta,
\]
then
\begin{multline}
      \label{eq-unt2}
\big|G_0(x,y;\Re z-h)- G_0(x,y;\Re z)\big|\le \dfrac{2\delta}{\pi} 
\sqrt[4]{\dfrac{\pi^2}{-4\Re z\,|x-y|^2}}\,  e^{-\sqrt{- \Re z}|x-y|}\\
\text{for } \Re z<-E(\delta,x,y), \quad |\Im z|<\gamma(\delta,x,y), \quad h\in\big(0,h_0(\delta,x,y)\big).
\end{multline}
Substituting now the estimates \eqref{eq-unt0}, \eqref{eq-unt}, and \eqref{eq-unt2} into \eqref{eq-gstart},
one shows that
\[
\big|G_h(x,y;z)\big|\ge \dfrac{1}{2\pi}\,\Big(\dfrac{1-\delta}{(1+\delta)^2}-\delta(1+\delta)-4\delta \Big)\,
\sqrt[4]{\dfrac{\pi^2}{-4\Re z\,|x-y|^2}}\,  e^{-\sqrt{- \Re z}|x-y|}
\]
for $\Re z<-E(\delta,x,y)$, $|\Im z|<\gamma(\delta,x,y)$, and $h\in\big(0,h_0(\delta,x,y)\big)$,
which yields the result due to the arbitrariness in the choice of $\delta$.
\end{proof}

\section{Estimates for the effective Hamiltonians}\label{sec5}

In this section, we are going to analyze the symbols $L_\alpha$
from Corollary \ref{corol2}. One can rewrite \eqref{eq-LL} as follows:
\begin{gather*}
L_\alpha(x,p;z,h)  =q(z,h)+\alpha^{-1} + a(z,h)\big(\cos x+\cos p
\big)+ W(x,p;z,h),\\
a(z,h):=2\lambda(1,0;z,h),\\
W(x,p;z,h):=\sum_{\substack{m,n\in\ZZ,\\|m|+|n|\ge 2}}\lambda(m,n;z,h)\, e^{i(mx+np)}\,.
\end{gather*}

First note that due to the integral representation, Eq. (13.2.5) in \cite{AS},
\[
\Gamma(a)U(a,c;x)=\int_0^\infty e^{-xt} t^{a-1} (1+t)^{c-a-1} dt,\quad \Re a>0,
\]
and to the formula \eqref{eq-green} for the Green function one has
\begin{multline*}
G_h(x,y;z)=\dfrac{1}{4\pi}\,\exp\Big(- \dfrac{i h (x_1-y_1)(x_2+y_2)}{2}-\dfrac{h(x-y)^2}{4}\Big)\\
\times\int_0^\infty
e^{-\tfrac{h(x-y)^2t}{2}}\, \Big(\dfrac{t}{t+1}\Big)^{\tfrac{h-z}{2h}}\,\dfrac{dt}{t}.
\end{multline*}
Therefore, by \eqref{eq-lambda}, there exists
 $C=C(h)>0$ such that for $\Re z<0$ one has
\[
\big|\lambda(m,n;z,h)\big|\le C \exp\Big(-\dfrac{h(m^2+n^2)}{4}\Big)
\text{ for all } m \text{ and }n,\, |m|+|n|>0.
\]
Hence $L_\alpha$ is an analytic function of $(x,p)\in\CC^2$.

Note that due to Corollary \ref{prop-zb} and \eqref{eq-gf0} one has
\[
\big|a(z,h)\big|\ge \dfrac{e^{-h/4}}{2\pi}\, K_0(\sqrt{h-z}), \quad
h>0, \quad z\in(-\infty,h). 
\]
As the function $K_0(w)$ is strictly positive for positive $w$, see
e.g. 
Section~9.6.1 in \cite{AS},
one can conclude that $a(z,h)$ is non-zero for all positive $h$ and real negative $z$,
and Corollary \ref{corol2} gives 
\begin{lem} \label{lem-a3}
The negative spectrum of $H_{h,\alpha}$
coincides with the negative $z$-spectrum of a family of $h$-pseudodifferential operators
associated with the symbols
\begin{multline}
       \label{eq-lm}
M_\alpha(x,p;z,h)=\dfrac{1}{a(z,h)}\,\, L_\alpha(x,p;z,h) \\
=m_\alpha(z,h)+\cos x+\cos p+ N(x,p;z,h)
\end{multline}
with
\begin{equation}
\label{defmalpha}
\begin{gathered} m_\alpha(z,h)=\dfrac{q(z,h)+\alpha^{-1}}{a(z,h)},\\
N(x,p;z,h)=\sum_{\substack{m,n\in\ZZ,\\|m|+|n|\ge 2}}\dfrac{\lambda(m,n;z,h)}{a(z,h)}\, e^{i(mx+np)}.
\end{gathered}
\end{equation}
\end{lem}

\begin{lem}\label{prop3}
Take $h_0>0$ and $R>0$. Then one can find $E_0>0$  with the following property\textup{:}
for any $E>E_0$ there exists $\alpha_E>0$ such that for
$\alpha\in(0,\alpha_E)$, $h\in(0,h_0)$, $\Re z<-E_0$
the condition $\big|\,q(z,h)+\alpha^{-1}\big|\le R \big|a(z,h)\big|$
guarantees  that $\Re z<-E$ and $|\Im z|<4h_0$.
\end{lem}

\begin{proof} 
Take an arbitrary $E'>0$. Due to Lemma \ref{lem-oben}, there exists $C'$ such that,
for $\Re  z<-E'$  and $h>0$,  we have the estimate
\begin{equation*}\label{estisura}
|a(z,h)|\le C' \exp\big(-\sqrt{-\Re z}\big)\;.
\end{equation*}
Therefore, the condition $$\big|q(z,h)+\alpha^{-1}\big|\le R \, \big|a(z,h)\big|$$ implies
\[
\big|q(z,h)+\alpha^{-1}\big|
\le C \exp(-\sqrt{-\Re z}\big),\quad C=C' R.
\]
Using the triangular inequality,
\[\big|q_0(z)+\alpha^{-1}\big| \leq 
\big|q(z,h)+\alpha^{-1}\big|  +
\big|q(z,h)-q_0(z)\big|,
\]
and  the estimate \eqref{eq-qh1} we obtain
\[
\big|q_0(z)+\alpha^{-1}\big|\le \dfrac{1}{4\pi}\log\dfrac{\Re z-h}{\Re z}+C e^{-\sqrt{-\Re z}}\le
\dfrac{h}{4\pi|\Re z|}+C e^{-\sqrt{|\Re z|}}.
\]
Substituting here the explicit expression \eqref{eq-qex} we obtain
\[
\Big|-\dfrac{\log|z| +i\arg (-z)-\log 4-2\psi(1)}{4\pi}
+\alpha^{-1}\Big|\le
\dfrac{h}{4\pi|\Re z|}+C e^{-\sqrt{|\Re z|}}.
\]
In particular, considering the real and the imaginary parts separately,
\begin{gather}
       \label{eq-qim1}
\Big|-\dfrac{\log|z|-\log 4-2\psi(1)}{4\pi}+\alpha^{-1}\Big| \le \dfrac{h}{4\pi|\Re z|} + C e^{-\sqrt{-\Re z}},\\
       \label{eq-qim2}
\dfrac{|\arg(-z)\big|}{4\pi} \le \dfrac{h}{4\pi|\Re z|} +
C e^{-\sqrt{-\Re z}}.
\end{gather}
Using \eqref{eq-qim2}, let us choose $E_0> E'$ such that
\begin{gather*}
\big|\arg(-z)\big|\le \dfrac{\pi}{4}, \quad \Re z<-E_0, \quad h\in(0,h_0),\\
\log 2E_0-\log 4-2\psi(1)>0.
\end{gather*}
Then, under  the same conditions on $z$ and $h$, one has 
\[
|\Re z|\le|z|\le 2|\Re z|\;,
\]
and from \eqref{eq-qim1} one can get
\begin{equation}
        \label{eq-qim3}
\alpha^{-1} \le \dfrac{h}{4\pi|\Re z|} + C e^{-\sqrt{-\Re z}}+\dfrac{\log 2|\Re z|-\log 4-2\psi(1)}{4\pi}.
\end{equation}
With 
\[
B:=\sup\Big\{
\dfrac{h}{4\pi|\Re z|} + C e^{-\sqrt{-\Re z}} -\dfrac{2\psi(1)+\log 4}{4\pi}: \, \Re z<-E_0, \, h\in(0,h_0)
\Big\}
\]
the estimate \eqref{eq-qim3} gives
\begin{equation}
        \label{eq-al1}
\alpha^{-1}\le B+\dfrac{\log 2|\Re z|}{4\pi}.
\end{equation}
Therefore, for any $E>E_0$ and $\alpha\in(0,\alpha_E)$ 
with
\[
\alpha_E=\dfrac{4\pi}{4\pi B+\log 2E}
\]
the condition \eqref{eq-al1} gives $\Re z<-E$.

Using \eqref{eq-qim2} again, one can (if necessary) increase $E_0$ to obtain
\[
\big|\arg(-z)\big|\le \dfrac{2 h_0}{|\Re z|}, \quad \Re z<-E_0,
\]
and
\[
\tan t \le 2t \text{ for } t\in\Big(0,\dfrac{2h_0}{E_0}\Big),
\]
then one has 
\[
|\Im z|= |\tan \arg(-z)| \cdot |\Re z|< 4h_0
\]
for $\Re z<-E_0$.
\end{proof}

\begin{lem}\label{prop4} Take  $h_0>0$ and $R>0$ and let $E_0$ be the corresponding constant from Lemma~\ref{prop3}.
There exists $\alpha_0>0$ such that, for any $h\in(0,h_0)$ and $\alpha\in(0,\alpha_0)\,,$
the equation
\begin{equation}
     \label{eq-iter1}
q(z,h)+\alpha^{-1}=\mu a(z,h)
\end{equation}
with respect to $z$, $\Re z<-E_0$, has a unique solution $z=\zeta_\alpha(\mu,h)$
holomorphic in $\mu$ for complex $\mu$ with $|\mu|<R$, and this solution
is real for real $\mu$.
\end{lem}

\begin{proof}
For complex $\beta$ and $\alpha>0$ denote by $z(\beta,\alpha)$ the solution $z$
to the equation
\[
q_0(z)+\alpha^{-1}=\beta.
\]
Due to the explicit formula \eqref{eq-q0} for $q_0$ one has
an explicit expression
\begin{equation}
     \label{eq-zbeta}
z (\beta,\alpha)  =  - \exp \left( \log 4 + 2 \psi(1) +
\frac{4\pi}{\alpha}- 4 \pi \beta \right);
\end{equation}
in particular,
\[
\Re z (\beta,\alpha) = - \exp \left( \log 4 + 2 \psi(1) +
\frac{4\pi}{\alpha}- 4 \pi\Re \beta \right)\,\cos (4 \pi \Im \beta).
\]

Now note that if $\beta=\beta(\alpha,\mu,h)$ satisfies
\[
\beta = q_0\big(z(\beta,\alpha)\big) - q\big(z(\beta,\alpha),h\big) + \mu a\big(z(\beta,\alpha),h\big),
\]
then $\zeta_\alpha(\mu,h):=z\big(\beta(\alpha,\mu,h),\alpha\big)$
is a requested solution for \eqref{eq-iter1}.

To show the existence of a unique $\beta(\alpha,\mu,h)$ we use the fixed point theorem.
We are going to analyze the map $\Psi$,
\[
\beta \mapsto \Psi (\beta,\alpha,\mu,h):= q_0\big(z(\beta,\alpha)\big) - q\big(z(\beta,\alpha),h)\big)
+ \mu a\big(z(\beta,\alpha),h\big),
\]
and to show that it is a contraction for $|\beta|\le 1/16$ if $\alpha$
 is sufficiently small.
Namely, we will prove that one can choose $\alpha_0$ in such a way
that
\begin{equation}
       \label{eq-iter2}
\big|\Psi(\beta,\alpha,\mu,h)
\big|\le \dfrac{1}{32} \text{ for }
|\beta|\le\dfrac{1}{16}, \, \alpha\in(0,\alpha_0), \,
|\mu|< R,\, h\in(0,h_0),
\end{equation}
and that
\begin{multline}
       \label{eq-iter3}
\Big|\Psi(\beta_1,\alpha,\mu,h)-\Psi(\beta_2,\alpha,\mu,h) \Big|\le \dfrac{|\beta_1-\beta_2|}{2}\\
\text{ for }
|\beta_1|,|\beta_2|\le\dfrac{1}{16}, \, \alpha\in(0,\alpha_0), \,
|\mu|< R,\, h\in(0,h_0),
\end{multline}
then, by the fixed point theorem, the equation $\beta=\Psi(\beta,\alpha,\mu,h)$
has a unique solution with $|\beta|\le 1/16$ depending holomorphically on $\mu$, which can be obtained
throught by iteration $\beta_{n+1}=\Phi(\beta_n,\alpha,\mu,h)$ with, say, $\beta_0=0$,
and taking the limit $\beta=\lim \beta_n$. If $\mu$ is real, then all $\beta_n$ are real as well,
hence so is $\beta$.

Note that for $|\beta|\le 1/16$ one has
\begin{equation}
       \label{eq-zba}
\Re z (\beta,\alpha)\le - \dfrac{1}{\sqrt 2}\,\exp \big( \log 4 + 2 \psi(1) +
\frac{4\pi}{\alpha}- 4 \pi\Re \beta \big).
\end{equation}

To show \eqref{eq-iter2} we note that, by Lemma \ref{Lemma8} and
Lemma \ref{lem-oben} there exists $E_1>0$ such that for $\Re z(\beta,\alpha)<-E_1$
and $|\mu|<R$ one has
\[
\big|\Psi(\beta,\alpha,\mu,h)\big|\le
\dfrac{h}{4\pi \big|\Re z (\beta,\alpha)\big|}+
\dfrac{R}{|\Re z(\beta,\alpha)|^{1/4}}\, \exp\big(-\sqrt{|\Re z(\beta,\alpha)|}\big).
\]
Therefore, there exists $E_2>E_1$ such that
\[
\big|\Psi(\beta,\alpha,\mu,h)\big|\le \dfrac{1}{32}
\text{ for } h\in(0,h_0),\, |\mu|<R,\,\Re z(\beta,\alpha)<-E_2,
\]
and it remains to use \eqref{eq-zba} to choose $\alpha_1>0$ in such a way
that $\Re z(\beta,\alpha)<-E_2$ for $\alpha\in(0,\alpha_1)$ and $|\beta|\le 1/16$.

Let us prove now the estimate \eqref{eq-iter3}.
One has
\[
\Big|\Psi(\beta_1,\alpha,\mu,h)-\Psi(\beta_2,\alpha,\mu,h) \Big|
\le \Phi(\alpha,\mu,h) |\beta_1-\beta_2|,
\]
where
\[
\Phi(\alpha,\mu,h):=\sup_{|\beta|\le 1/16}
\Big|
\dfrac{\partial \Psi(\beta,\alpha,\mu,h)}{\partial \beta}
\Big|.
\]
Hence it is sufficient to show that
\begin{equation}
      \label{eq-partf}
\Big|
\dfrac{\partial \Psi(\beta,\alpha,\mu,h)}{\partial \beta}
\Big|\le 1/2
\end{equation}
for $|\beta|\le 1/16$ and  small $\alpha$.

We note first that, by \eqref{eq-zbeta}
and \eqref{eq-zba},
for $|\beta|\le 1/16$ there exist $B,C>0$ such that
\begin{equation}
   \label{eq-fbb}
\Big|\dfrac{\partial z(\beta,\alpha)}{\partial \beta}\Big|\le
B \exp \dfrac{4\pi}{\alpha} \le C \big|\Re z(\beta,\alpha)\big|.
\end{equation}
By Lemma \ref{Lemma8} and Lemma \ref{lem-afh}, there exists
$E_3>0$ such that for $\Re z(\beta,\alpha)<-E_3$ one has
\begin{multline*}
\Big|
\dfrac{\partial \Psi(\beta,\alpha,\mu,h)}{\partial \beta}
\Big|
\\
 \le 
\Big|\dfrac{\partial z(\beta,\alpha)}{\partial \beta}\Big|\,\bigg(
\dfrac{h}{4\pi \big|\Re z(\beta,\alpha)\big|^2}
+\dfrac{R}{\big|\Re z(\beta,\alpha)\big|^{3/4}} e^{-\sqrt{|\Re z(\beta,\alpha)|}}
\bigg)\\
\le
\dfrac{Ch }{4\pi \big|\Re z(\beta,\alpha)\big|}
+CR\big|\Re z(\beta,\alpha)\big|^{1/4} e^{-\sqrt{|\Re z(\beta,\alpha)|}},
\end{multline*}
where the last estimate is due to \eqref{eq-fbb}.
Hence one can find $E_4>E_3$ such that \eqref{eq-partf} holds 
provided $\Re z(\beta,\alpha)<-E_4$ for $|\beta|\le 1/16$ and $h\in(0,h_0)$,
and one can again pick $\alpha_2\in(0,\alpha_1)$ in such a way
that $\Re z(\beta,\alpha)<-E_4$ for $|\beta|\le 1/16$ for $\alpha\in(0,\alpha_2)$
using the estimate \eqref{eq-zba}.

Now it remains to show that for sufficiently small $\alpha$ any solution $z$ to \eqref{eq-iter1}
really satisfies
$\big|q_0(z)+\alpha^{-1}\big|\le 1/16$;
this will mean that the above solution $\zeta_\alpha$ is unique.
For any such solution $z$ one has $|q(z,h)+\alpha^{-1}|\le R |a(z,h)|$.
Combining this with the triangular inequality and the estimate of Lemma \ref{Lemma8}
one arrives at
\[
\big|q_0(z)+\alpha^{-1}\big|\le
\big|q(z,h)+\alpha^{-1}\big| +\big|q(z,h)-q_0(z)\big|\\
\le R \,|a(z,h)| +\dfrac{h}{4\pi|\Re z|}.
\]
By Lemma \ref{prop3}, for any $E>E_0$ we can find $\alpha_E\in(0,\alpha_2)$
such that for $\alpha\in(0,\alpha_E)$ and $h\in(0,h_0)$
the solution satisfies $\Re z<-E$. Take $E$ sufficiently large
to have
\[
R |a(z,h)| +\dfrac{h_0}{4\pi|\Re z|} \le 1/16 \text{ for } \Re z<-E,
\]
which is possible due to the estimate of Lemma \ref{lem-afh} for $a(z,h)$.
Denoting the corresponding value of $\alpha_E$ by $\alpha_0$, 
this completes the proof.
\end{proof}

Below we will need the fact that $a(z,h)$ does not vanish in a certain tubular complex domain,
and not only on the negative halfline. Namely, it follows from Lemma \ref{lem-unten} that one can find $h_*>0$
and $E_*>0$ such that~:
\begin{equation*}\label{proprietea}
 a(z,h)\neq 0\;,\; \mbox{ if } \Re z<-E_*\,,\, |\Im z|<4 h_* \mbox{
   and } h\in(0,h_*)\,.
\end{equation*}
Note that in this domain the equation \eqref{eq-iter1} reads simply as $m_\alpha(z,h)=\mu$.

The following theorem summarizes all previous considerations and shows the applicability
of the Helffer-Sj\"ostrand methods~\cite{HS1,HS2,HS3}.

\begin{theorem}\label{th10} For any $E_0>0$
there exist $\alpha_0>0$ and $h_0>0$ such that for
$\alpha\in(0,\alpha_0)$ and $h\in (0,h_0)$ one has
\[
\spec H_{h,\alpha}\cap(-\infty,-E_0)=\big\{
z<-E_0: m_\alpha(z,h)\in\mspec \Hat P_{h,\alpha}(\mu)
\big\},
\]
where $m_\alpha$ is given in \eqref{defmalpha},
 $\Hat P_{h,\alpha}$ is a pseudodifferential operator with a strong type I symbol
$P_\alpha(x,p;\mu,h)$  given by Eq.~\eqref{eq-pm} below, and $\lim_{\alpha\to 0} \varepsilon(P_\alpha)=0$.
\end{theorem}

\begin{proof} By Corollary \ref{corol2} and Eq. \eqref{eq-lm},
the negative spectrum of $H_{h,\alpha}$
coincides with the $z$-spectrum for the symbols $M_\alpha(x,p;z,h)$.
Note that the symmetry conditions (c) in Definition~\ref{defin1}
are satisfied due to the specific form of the Fourier coefficients, see \eqref{eq-lambda},
so we will be concerned below with the conditions (a) and (b) of Definition~\ref{defin1}.

Due to Lemma \ref{lem-oben} and the estimate \eqref{eq-qh1}, for any segment $[a,b]\subset(-\infty,0)$
there exists a constant $C>0$ such that $\|Q(z,h)\|\le C$ for $z\in[a,b]$ and any $h$.
By \eqref{eq-specba}, this means that $H_{h,\alpha}$ has no spectrum in $[a,b]$
for $|\alpha|<1/C$. Therefore, we can assume without loss of generality that $E_0$
satisfies the assumptions of Lemma \ref{prop3} for $R=5$ and $h_0=h_*$, and that $E_0\ge E_*$.

Using Lemma \ref{prop4} one can choose $\alpha_1>0$
such that  for $h\in(0,h_*)$ and $\alpha\in(0,\alpha_1)$
the equation $m_\alpha(z,h)=\mu$ with respect to $z$ with $\Re z<-E_0$
has a unique solution $z=\zeta_\alpha(\mu,h)$ for all $\mu$ with $|\mu|\le 4$,
and this solution is a holomorphic function of $\mu$ and is real for real $\mu$.

Let us consider the symbols 
\begin{equation}
    \label{eq-pm}
P_\alpha(x,p;\mu,h):=M_\alpha(x,p;\zeta_\alpha(\mu,h),h).
\end{equation}
Clearly, 
\[
P_\alpha(x,p;\mu,h):=\mu +\cos x+\cos p+T_\alpha(x,p;\mu,h),
\]
where
\begin{equation*}
T_\alpha(x,p;\mu,h):=\sum_{\substack{m,n\in\ZZ,\\|m|+|n|\ge 2}}t_\alpha(m,n;\mu,h) e^{i(mx+np)}\end{equation*}
and
\begin{equation*}
t_\alpha(m,n;\mu,h):=\dfrac{\lambda\big(m,n;\zeta_\alpha(\mu,h),h\big)}{a\big(\zeta_\alpha(\mu,h),h\big)}.
\end{equation*}
As previously noted, the symbols $M_\alpha$ and hence $T_\alpha$ are analytic
with respect to $x$ and $p$ in the whole complex space.
Furthermore, for real $\mu$ the solutions $\zeta_\alpha(\mu,h)$ are also real, hence,
by \eqref{eq-lambda}, the coeffcieints $t_\alpha(m,n;\mu,h)$ are real as well and,
as noted above, satisfy $t_\alpha(m,n;\mu,h)\equiv t_\alpha(-m,-n;\mu,h)$. Therefore,
$P_\alpha$ takes only real values for real $\mu$.

Now take an arbitrary $\varepsilon>0$. By Lemmas \ref{lem-oben} and \ref{lem-unten}
there exists $E_1>E_0$ and $h_0\in(0,h_*)$ such that
\[
\big|t_\alpha(m,n;\mu,h)\big|\le 2\exp\Big(-\sqrt{-\Re \zeta_\alpha(\mu,h)}\,\big(\sqrt{m^2+n^2}-1\big)\Big).
\]
for $\Re \zeta_\alpha(\mu,h)<-E_1$, $\big|\Im\zeta_\alpha(\mu,h)\big|<4h_0$,  $h\in(0,h_0)$, and
$|m|+|n|>1$. Then one can estimate for $|\Im x|+|\Im p|<\varepsilon^{-1}$
(hence for $|\Im x|<\varepsilon^{-1}$ and $|\Im p|<\varepsilon^{-1}$):
\[
\big|T_\alpha(x,p;\mu,h)\big|\le 2 \sum_{\substack{m,n\in\ZZ,\\|m|+|n|\ge 2}}
\exp\Big(\dfrac{|m|+|n|}{\varepsilon}-\sqrt{-\Re\zeta_\alpha(\mu,h)}\,\big(\sqrt{m^2+n^2}-1\big)\Big).
\]
Therefore, there exists $E>E_1$ such that
\[
\big|T_\alpha(x,p;\mu,h)\big|\le\varepsilon \text{ for }
 |\Im x|+|\Im p|<\dfrac{1}{\varepsilon}
\]
provided
\[
|\mu|\le 4, \quad \Re \zeta_\alpha(\mu,h)<-E, \quad \big|\Im\zeta_\alpha(\mu,h)\big|<4h_0, \quad  h\in(0,h_0).
\]
Now using Lemma \ref{prop3} we can choose $\alpha_0\in(0,\alpha_1)$ such that
$\Re \zeta_\alpha(\mu,h)<-E$ and $\big|\Im\zeta_\alpha(\mu,h)\big|<4h_0$ for all $\alpha\in(0,\alpha_0)$, $h\in(0,h_0)$, and $|\mu|\le 4$.
This means that for all $\alpha\in(0,\alpha_0)$ and $h\in(0,h_0)$ the symbol $P_\alpha(\mu,h)$
is of strong type I with $\varepsilon(P_\alpha)\le\varepsilon$.
\end{proof}

Using Theorem \ref{ths}, one can get the following corollary.

\begin{corol} For any $E_0>0$ there exist $\alpha_0>0$ and $C>0$ such that for $\alpha\in(0,\alpha_0)$
and  sufficiently small irrational
$(2\pi)^{-1}h$ admitting a continuous fraction expansion 
\[
\dfrac{h}{2\pi}=\dfrac{1}{n_1+\dfrac{\mathstrut 1}{n_2+\dfrac{\mathstrut 1}{n_3+\dots}}}
\]
with integers $n_j$ satisfying $n_j\ge C$, the spectrum of $H_{h,\alpha}$ in $(-\infty,-E_0)$
is a zero measure Cantor set.
\end{corol}

\begin{proof}
For sufficiently small $\varepsilon(P_\alpha)$ and $h$ as described, the set 
 $\mspec \Hat P_{h,\alpha}(\mu)$ is a zero measure Cantor set in virtue of Theorem~\ref{ths}. In view of Theorem \ref{th10}
 it remains to emphasize that $z\mapsto m_\alpha(z,h)$ is an analytic topological isomorphism
 for sufficiently small $\alpha$ and $h$ (see Lemma \ref{prop4}), hence
 the preimage of a zero-measure Cantor set is a zero-measure Cantor as well.
\end{proof}

\begin{remark}\normalfont
In \cite{BGP} another solvable model was considered, namely,
periodic quantum graphs with magnetic fields, and the
expression for the spectrum obtained there in Theorem~13 is very close to the one of Theorem~\ref{th10}.
Nevertheless, the essential difference was that the term $T_\alpha$ in the constructions
 analogous  to Theorem~\ref{th10} would just vanish,
hence one had a reduction to the usual spectral problem for the discrete
magnetic Laplacian. It is worthwhile to note that, even with such a simplification,
the spectrum of non-isotropic quantum graphs was studied only numerically \cite{gold} up to now.
\end{remark}

\section*{Acknowledgments}
The second named author gratefully acknowledges a partial support by the Deutsche Forschungsgemeinschaft
and by a  fellowship of the French Government.

\end{document}